\documentclass[sn-mathphys,Numbered]{sn-jnl}

\usepackage{graphicx}%
\usepackage{multirow}%
\usepackage{amsmath,amssymb,amsfonts}%
\usepackage{amsthm}%
\usepackage{mathrsfs}%
\usepackage[title]{appendix}%
\usepackage{xcolor}%
\usepackage{textcomp}%
\usepackage{manyfoot}%
\usepackage{booktabs}%
\usepackage{algorithm}%
\usepackage{algorithmicx}%
\usepackage{algpseudocode}%
\usepackage{listings}%
\usepackage{bm}

\theoremstyle{thmstyleone}%
\theoremstyle{thmstyletwo}%

\theoremstyle{thmstylethree}%

\begin{document}

\title[About the performance of perturbative treatments of the spin-boson dynamics within the hierarchical equations of motion approach]{About the performance of perturbative treatments of the spin-boson dynamics within the hierarchical equations of motion approach}


\author*[1]{\fnm{Meng} \sur{Xu}}\email{meng.xu@uni-ulm.de}
\author*[1]{\fnm{Joachim} \sur{Ankerhold}}\email{joachim.ankerhold@uni-ulm.de}
\affil*[1]{Institute  for Complex Quantum Systems and IQST, Ulm University, \\ Albert-Einstein-Allee 11, D-89069  Ulm, Germany}

\abstract{The hierarchical equations of motion (HEOM) provide a numerically exact approach for simulating the dynamics of open quantum systems coupled to a harmonic bath. However, its applicability has traditionally been limited to specific spectral forms and relatively high temperatures. Recently, an extended version called Free-Pole HEOM (FP-HEOM) has been developed to overcome these limitations. In this study, we demonstrate that the FP-HEOM method can be systematically employed to investigate higher-order master equations by truncating the FP-HEOM hierarchy at a desired tier. We focus on the challenging scenario of the spin-boson problem with a sub-Ohmic spectral distribution at zero temperature and analyze the performance of the corresponding master equations. Furthermore, we compare the memory kernel for population dynamics obtained from the exact FP-HEOM dynamics with that of the approximate NIBA (Non-Interacting-Blip Approximation).}




\maketitle

\section{Introduction}
The second-order quantum master equation (QME) is a fundamental tool for studying the dynamics of open quantum systems and finds wide-ranging applications in diverse fields, including quantum optics, condensed matter physics, condensed phase chemistry, and biology \cite{gardiner2004quantum,scully1997quantum,kamenev2023field,nitzan06,may11,mohseni2014quantum}. However, employing the QME in regimes with strong non-Markovian behavior, characterized by slow environmental evolution and significant retardation effects on the subsystem, presents substantial challenges. These challenges become even more pronounced in low-temperature conditions and structured environments, where non-Markovian effects are amplified. A notable example is the unconventional impact of sub-Ohmic noise at zero temperature \cite{weiss12,kast2013persistence,xu2022taming}.

To address the limitations of the QME in handling pronounced non-Markovian retardation effects, incorporating higher-order corrections that account for system-bath correlations becomes essential \cite{zhang16,xu17,xu2018convergence}. However, direct calculations involving these terms require computing high-dimensional time integrals, which is a daunting task \cite{shibata80,terwiel74}. To the best of our knowledge, analytical calculations of high-order perturbations or kernels have only been achieved up to the sixth order \cite{gong15,aihara90}. Previous work has employed numerical techniques such as hierarchical equations of motion (HEOM) to compute higher-order effects \cite{xu17,xu2018convergence}. However, these approaches have been primarily limited to elevated-temperature regimes due to the exponential increase in the number of auxiliary density operators (ADOs) involved, a challenge commonly referred to as the curse of dimensionality. This challenge becomes even more severe as the temperature approaches zero, making the analysis of memory kernels' properties and high-order perturbative master equations increasingly demanding. 

To investigate the effects of high-order kernels in unconventional environments, we address key challenges such as zero-temperature calculations employing the widely used HEOM method \cite{tanimura89,tanimura06,tanimura2020numerically,zhang2020hierarchical,hsieh2018unified,wang2019dynamical,shi09b,liu14,xu17,shi2018efficient,han2019stochastic,yan2019unified,erpenbeck2018extending,rahman2019chebyshev,yan2020new,ikeda2020generalization,nakamura18,hartle13,dunn2019removing,yan2016stochastic,xu2022taming,ke2022hierarchical,li2023dissipatons,yan2021efficient,li2022low,ke2023tree,chen2022universal,fay2022simple}. The HEOM approach utilizes a set of ADOs to unravel system-bath correlations within an extended state space \cite{yan14a,li2022low,liu14,ikeda2020generalization,yan2020new}. Conventionally, these ADOs are constructed as high-dimensional arrays based on a series of exponential functions derived from the decomposition of the bath correlation function. The extended space, expanded by ADOs, grows exponentially with the number of modes, posing significant challenges in scenarios involving low temperatures and structured bath spectra. This is particularly relevant as standard analytical Matsubara frequencies (i.e., $\nu_k = 2k\pi/\beta$) shift towards a continuum strip as temperatures decrease ($\beta = 1/T$). Therefore, developing an efficient method for obtaining a limited number of well-behaved modes becomes crucial for successfully implementing HEOM in the study of quantum master equations.

Several decomposition schemes \cite{chen2022universal,dan2022efficient,xu2022taming} have been developed to address the limitations of HEOM in low-temperature and general spectral density scenarios, such as the frequency-domain barycentric spectrum decomposition (BSD) \cite{xu2022taming}. The effectiveness of these decomposition schemes has been demonstrated by their ability to reproduce phenomena like the Kondo resonance peak in Fermi baths \cite{dan2022efficient} and the Shiba relation in sub-Ohmic boson baths \cite{xu2022taming}. Notably, by optimizing the properties of auxiliary modes under specific constraints, these schemes allow for the integration of a minimal number of modes (poles) into various versions of HEOM and other methods, such as the pseudomode approach \cite{tamascelli2018nonperturbative}, the hierarchy of pure states (HOPS) \cite{suess14}, and the nonequilibrium dynamical mean-field theory (DMFT) \cite{arrigoni2013nonequilibrium}, even in unconventional environments.

In this work, we employ the Free-Poles HEOM (FP-HEOM) \cite{xu2022taming}, which combines the optimized BSD and a hierarchical structure, to showcase the performance of higher-order master equations in addressing unconventional environments. We also compare it with the well-established Non-Interacting-Blip Approximation (NIBA), a powerful but approximate scheme for non-perturbative treatment of the system-bath coupling in spin systems. The remainder of this paper is organized as follows: In Sections II and III, we provide a brief overview of the hierarchical equations of motion and discuss their limitations in simulating dynamics of open quantum systems. In Section IV, we present numerical simulation results to demonstrate the efficiency and accuracy of our proposed method. Finally, we summarize our findings and discuss potential avenues for future research in the concluding section.

\section{Model Hamiltonian}
The spin-boson model, which describes a two-state system interacting bilinearly with a harmonic bath, has been extensively studied in open quantum system dynamics\cite{weiss12}. In this work, we employ this model as a prototypical example to benchmark our method, while generalizations to more complex models are straightforward.

We consider a spin-boson model with the Hamiltonian
\begin{equation}\label{Eq:htot}
\begin{split}
H_T =& \epsilon\sigma_z + \Delta\sigma_x + \sum_ic_ix_i\sigma_z \\
& + \sum_i{\frac{p_i^2}{2m_i} + \frac{1}{2}m_i\omega_i^2x_i^2} + \frac{1}{2}\mu\sigma_z^2 \;\;,
\end{split}
\end{equation}
where the system degrees of freedom (DoF) are represented by dimensionless Pauli matrices $\sigma_x$ and $\sigma_z$. $\Delta$ and $2\epsilon$ denote the tunneling energy and energy bias between the two eigenstates of $\sigma_z$, i.e., {$|\pm\rangle$}. The $i$th harmonic modes of the reservoir is characterized by its mass, coordinate, momentum, and frequency, i.e.\ $m_i$, $x_i$, $p_i$, and $\omega_i$, respectively. The coupling between the system and the $i$th bath mode is denoted by $c_i$ with $[m\omega^2x]$ dimension.

The effective impact of the bath on the system is fully described by the coupling-weighted spectral density
\begin{equation}
J(\omega) = \frac{\pi}{2}
\sum_{i} \frac{c_i^2}{m_i\omega_i^2}
\delta(\omega-\omega_i) \;\; .
\end{equation}
In the following, we assume a generic spectral density \cite{leggett87} of the form
\begin{equation}\label{Eq:general}
J(\omega) = \frac{\pi}{2}\,\alpha\, \omega_c^{1-s}\,\omega^{s}\, \mathrm{e}^{-\omega/\omega_c} \;\;,
\end{equation}
where $\alpha$ is the dimensionless Kondo parameter characterizing the system-bath dissipation strength and $\omega_c$ is the characteristic frequency of the bath. For $s=1$ this spectral distributions describes an ohmic reservoir and it is calledf sub-ohmic for $s<1$. In the latter situation, 
the interplay of the relatively large portion of  low frequency modes with the internal two level dynamics makes this model extremely challenging to simulate, particularly in the long time limit and close to or at $T = 0$. 

\section{Hierarchical equations of motion}
\label{sec:heom}
We briefly describe the essence of the HEOM approach and refer to the literature for further details.
The derivation assumes a factorized initial states of the total density at time zero, i.e.\ $W(0) = \rho_s(0)\otimes e^{-\beta H_b}/\rm{Tr}\, e^{-\beta H_b}$. The generalization to correlated initial states can be found in Refs.~\cite{tanimura14,song15b}.

In path integral representation \cite{feynman63}, a formally exact expression for the reduced density is obtained by integrating over the bath degrees of freedom.
The effective impact of the bath onto the system dynamics is captured by the Feynman-Vernon influence functional \cite{feynman63} which reads
\begin{multline}\label{Eq:fv-if}
  \mathcal{F}[q^{+},q^{-}] = 
 -\int_{0}^{t}ds\int_0^sd\tau [q^{+}(s)-q^{-}(s)] \\
  \times\left[C(s-\tau)q^{+}(\tau) - C^{\ast}(s-\tau)q^{-}(\tau) \right] \;\; ,
\end{multline}
where $q^\pm(\tau)$ denote forward and backward paths with respect to the eigenstates $|\pm\rangle$. 
The influence functional describes arbitrary long-ranged self-interactions in time of the system paths determined by the bath auto-correlation $C(t)=\langle \xi(t) \xi(0)\rangle$ with the collective degree of freedom $\xi=\sum_i c_i x_i$. Particularly at low temperatures correlation functions are known to decay algebraically so that direct simulations of the path integral, e.g., via path integral Monte-Carlo (PIMC) \cite{muhlbacher2004path,kast2013persistence} are plagued by a degrading signal to noise ratio. 

The hierarchical equations of motion (HEOM) approach deals with this problem by converting the time-non-local path integral into a set of time-local differential equations through the introduction of auxiliary density operators (ADOs). The basic ingredient is a decomposition of $C(t)$ in a series of $K$ exponential modes, where $K$ has to be sufficiently bounded for the nested hierarchy to be practically solvable. To formulate a minimal set of modes is thus the prerequisite to access low temperatures and long times. As a matter of fact, this problem has been solved by us only recently  \cite{xu2022taming} by implementing the barycentric representation of rational functions for the spectral noise power $S_\beta(\omega)$, i.e.\ the Fourier transform of $C(t)$
\begin{equation}\label{Eq:bsd}
\begin{split}
    C(t) &= \frac{1}{\pi} \int_{-\infty}^{+\infty} d\omega\, S_{\beta}(\omega) \,\mathrm{e}^{-i\omega t} \\
        &= \sum_{k=1}^{K} \, d_k\, \mathrm{e} ^{-z_k t} + \delta C(t) ~~ t \geq 0 \;\;
\end{split}    
\end{equation}
with complex-valued amplitudes $d_k$ and frequencies $z_k = \gamma_k + i\omega_k$, $\gamma_k > 0$. Here, $S_\beta(-\omega$ and $S_\beta(\omega)$ are related by the fluctuation dissipation theorem which can be cast into the form $S_\beta(\omega) = 2[n_\beta(\omega) +1 ] J(\omega)$ with the Bose distribution $n_\beta(\omega) = 1/[\exp(\beta\omega) -1]$. The barycentric representation constructs a function $\tilde{S}_\beta(z)$ in the complex plane  such that it is along the real axis  an approximant of $S_\beta(\omega)$ (AAA algorithm, see \cite{nakatsukasa2018aaa}). The poles and the residues of $\tilde{S}_\beta$ appear as $\{d_k\}$ and $\{z_k\}$ in (\ref{Eq:bsd}). For details of this Free Pole-HEOM see \cite{xu2022taming}. One can show that the FP-HEOM thus operates with a minimal set of $K$ modes with the correction $\delta C(t)$  below a chosen threshold. 

The exponential function's self-derivative property in (\ref{Eq:bsd}) then allows to 'unravel' the time non-locality of the original Feynman Vernon path integral by introducing ADOs $\rho_{\bf m,n}(t)$ with multi-indices $\mathbf{n}=(n_1, \ldots, n_K)$ and $\mathbf{m}=(m_1, \ldots, m_K)$.
This then leads to a nested hierarchy of time-local evolution equations for the ADOs, i.e., 
\begin{multline}\label{Eq:FP-heom}
\dot{\hat{\rho}}_{\bf m,n} =
 -(i\mathcal{L}_s +\sum_{k=1}^{K} m_{k} z_{k} + \sum_{k=1}^{K} n_{k} z_{k}^{*} ) \hat{\rho}_{{\bf m,n}} -i\sum_{k=1}^{K} \sqrt{(m_k+1)\, d_k} \left[\hat{q},\hat{\rho}_{{\bf m}_k^{+},{\bf n}} \right] \\
 -i\sum_{k=1}^{K} \sqrt{(n_k+1)\, d_k^*}\left[\hat{q},\hat{\rho}_{{\bf m,n}_k^{+}} \right] -i\sum_{k=1}^{K} \sqrt{m_k\, d_k}\;\hat{q}\;\hat{\rho}_{{\bf m}_k^{-},{\bf n}} +i\sum_{k=1}^{K} \sqrt{n_k\, d_k^*}\;\hat{\rho}_{{\bf m,n}_k^{-}}\hat{q}
\;\;.
\end{multline}
The subscript $\bm{m}_k^{\pm}$ and $\bm{n}_k^{\pm}$ denote $\{m_1,\ldots,m_k\pm1,\ldots\,m_K\}$ and $\{n_1,\ldots,n_k\pm1,\ldots,n_K\}$, respectively.  The bare system evolution is propagated by $\mathcal{L}_s\rho = [H_s,\rho]$. Eventually, the physical reduced density matrix $\hat{\rho}_s$ corresponds to the multi-index  $\bm{m}=\bm{n}=0$. 
To boost the numerical efficiency of the FP-HEOM (\ref{Eq:FP-heom}), matrix product state (MPS) representations can be conveniently implemented which allows to tackle also asymptotic times down to temperature zero \cite{xu2022taming}.

Note that the truncation of the nested hierarchy of ADOs after  tier $L$ such that
\begin{equation}
    \sum_k (n_k+m_k)\leq L\, ,
\end{equation}
formally implies to include system-bath coupling strength up to order $\alpha^{2L}$\cite{xu05}. In fact, as we have shown previously \cite{xu2021heat,xu2022minimal},  a truncation at $L=1$ leads to a generalized Redfield equation. 

\section{Numerical Results}

In this section, we utilize the nonperturbative Free-Pole HEOM (FP-HEOM) method to investigate the dynamics of a two-level system coupled to a sub-Ohmic bosonic bath at zero temperature ($T=0$), characterized by a spectral density of the form $J(\omega)\propto \omega^s, 0\leq s\leq 1$. Our analysis focuses on two main aspects: (i) the convergence properties of higher-order master equations with respect to the system-bath coupling strength $\alpha$ and spectral exponent $s$, and (ii) the nonperturbative memory kernels, which serve as effective generators for the system dynamics. 

To ensure the accuracy of our calculations, we employ a barycentric decomposition with a tolerance threshold of $\delta C(t) \leq 10^{-3}$. The FP-HEOM is propagated using the time-dependent variational principle (TDVP) \cite{lubich15} with a matrix product states (MPS) representation, utilizing a maximum bond dimension of $\chi = 30$. The simulations are performed on a single Intel Xeon Gold 6252 CPU @ 2.1 GHz core, with the computation time ranging from a few minutes to several hours, depending on the specific scenario under investigation.

\subsection{Performance of Truncated Time Evolution Equations within FP-HEOM}

Various approximate treatments for open quantum dynamics have been developed, the most prominent ones based on considering the system-reservoir coupling up to second order. When a time scale separation exists between the rapidly decaying reservoir correlations and the relaxation dynamics of the reduced density operator, these methods lead to the Redfield master equation \cite{breuer02}. As mentioned above, an extended Redfield equation, referred to as Redfield-plus, appears naturally if the FP-HEIOM is limited to ADOs with $\sum_k (n_k+m_k) \le 1$. Namely, in the interaction picture, the ADOs of the zeroth tier (reduced density matrix) read
\begin{align}\label{Eq:0th-ados}  
 \dot{\hat{\rho}}^{I}(t) &= -i\sum_k[\hat{q}^{I}(t),\sqrt{d_k} \rho_{{\bf 0}_k^{+}, \bf 0}^{I}(t) + \sqrt{d_k^\ast} \rho_{{\bf 0},{\bf 0}_k^{+}}^{I}(t)] \,
 \end{align}
with the first-tier ADOs determined via
\begin{subequations}\label{Eq:1th-ados}
 \begin{align}
  \frac{d}{dt}\hat{\rho}_{{\bf 0}_k^{+},\bf 0}^{I}(t) &= -z_k\rho_{{\bf 0}_k^{+},\bf 0}^{I}(t) -i\sqrt{d_k} \hat{q}^I(t) \hat{\rho}_{\bf 0,0}^I(t) \;\;; \\
\frac{d}{dt}\hat{\rho}_{{\bf 0,0}_k^{+}}^{I}(t) &= -z_k^\ast\rho_{{\bf 0},{\bf 0}_k^{+}}^{I}(t) -i\sqrt{d_k^\ast} \hat{\rho}_{\bf 0,0}^I(t) \hat{q}^I(t) \;\;.
\end{align}  
\end{subequations}
Here, the index $I$ indicates the interaction picture. It is worth noting that the structure of Eqs. (\ref{Eq:0th-ados}) and (\ref{Eq:1th-ados}) resembles the one known from generalized Floquet theory for driven systems \cite{traversa2013generalized,magazzu2017asymptotic,magazzu2018asymptotic}.
The factorized initial state gives a  boundary condition with $\hat{\rho}_{\bf 0,0}(0) = \hat{\rho}(0)$ and all ADOs with $L>0$ are set to zero. 

The inhomogeneous differential equations in Eqs. (\ref{Eq:1th-ados}) can be formally solved  
\begin{subequations}
    \begin{align}
        \hat{\rho}_{\bm{0}_k^+,\bm{0}}^I(t) &= -i\sqrt{d_k} \int_0^t d\tau\, \mathrm{e}^{-z_k (t-\tau)} \hat{q}^I(\tau) \hat{\rho}_{\bf 0,0}^I(\tau) \;\;; \\ 
        \hat{\rho}_{\bm{0},\bm{0}_k^+}^I(t) &= -i\sqrt{d_k^\ast} \int_0^t\, d\tau \mathrm{e}^{-z_k^\ast (t-\tau)}  \hat{\rho}_{\bf 0,0}^I(\tau) \hat{q}^I(\tau) \;\;.  
    \end{align}
\end{subequations}
and then plugged in into Eq. (\ref{Eq:0th-ados}). Summation over the reservoir modes as in Eq. (\ref{Eq:bsd}) then leads to the following time-evolution equation in Born approximation
\begin{multline}
\label{eq:redfieldplus}
\frac{d}{dt}\hat{\rho}^{I}(t)
=-\int_{0}^{t}d\tau \, 
[\hat{q}^{I}(t),C(t-\tau) \hat{q}^{I}(\tau)\rho^{I}(\tau) ]  \\
-[\hat{q}^I(t), C^{*}(t-\tau)\rho^I(\tau)\hat{q}^I(\tau)] \;\;.
\end{multline}
Note that no Markovian coarse-graining in time is done here, not even on the level of the reduced density so that the reservoir induced retardation is fully taken into account in this order of system-bath coupling. One regains the conventional Redfield equation by setting $\rho^I(\tau)\to \rho^I(t)$, thus leading to a time-local evolution equation with time-dependent rates. Hence, we name the above integro-differential equation (\ref{eq:redfieldplus}) {\em Redfield-plus}.

Thus, the FP-HEOM can be viewed as an infinite-order extension of the Redfield-plus/Redfield approximation \cite{xu05,trushechkin2019higher}, where the truncation at tier $L$ 
yields an evolution equation non-local in time of order $\alpha^{2L}$. This enables a systematic analysis of the role of higher-order system-reservoir correlations, which are particularly subtle within perturbative theory.

\begin{figure}[htbp]
\centering
\includegraphics[width=8.5cm]{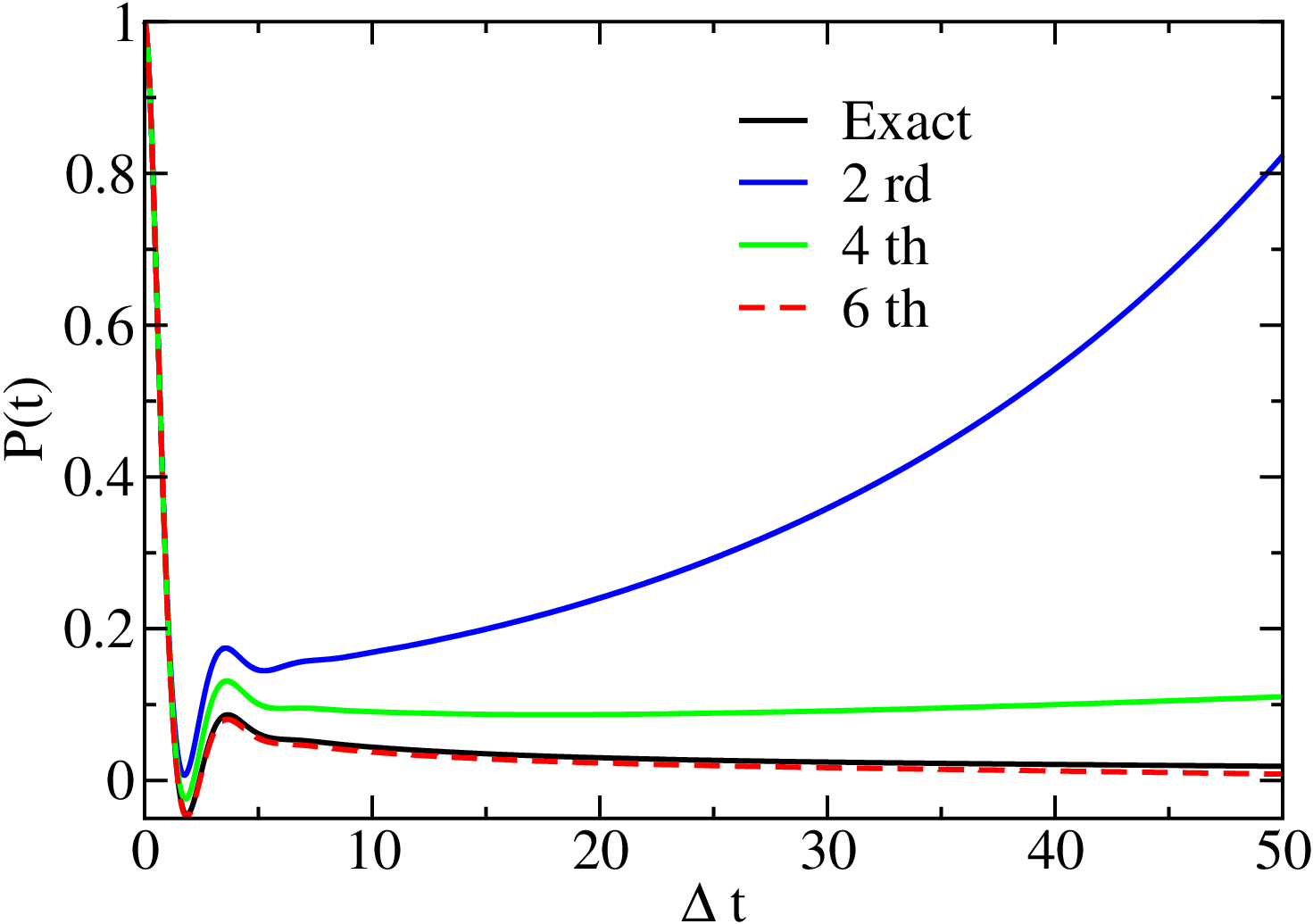}
\caption{High order perturbative time evolution equation according to tier $L$ of the FP-HEOM versus the fully converged simulations. For a sub-ohmic spin boson model the parameters are: $\epsilon = 0$, $\Delta = 1$, $\omega_c = 20$, $T = 0$, $\alpha = 0.1$, and $s = 0.5$.}
\label{f1}
\end{figure}

As demonstrated in Fig.~\ref{f1} with a coupling strength of $\alpha = 0.1$, numerically converged "exact" results can be achieved when the truncation tier reaches $L = 12$. However, the spin dynamics can be  approximated already quite decently in the $6$th order ($L = 3$), while the Redfield-plus ($L=1$) approximation has the tendency to diverge over longer timescales.
We remark, that while the FP-HEOM can treat also strong system-bath couplings, here, we have chosen a sufficiently weak parameter $\alpha$ to be in line with the concept of a perturbation series. 

\subsection{Redfield-plus: Influence of Low-Frequency Modes}

\begin{figure}[htbp]
\centering
\includegraphics[width=8.5cm]{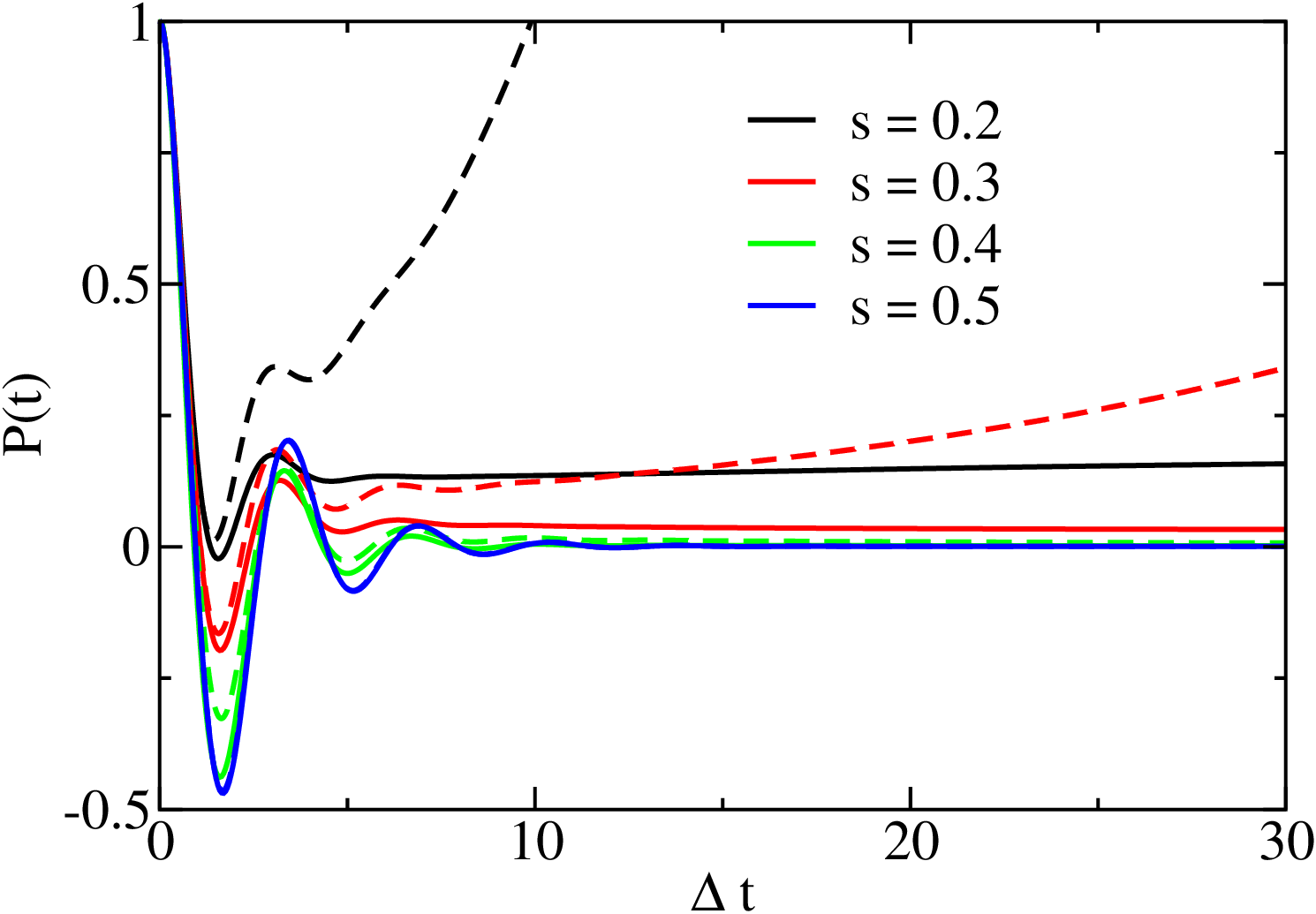}
\caption{Redfield-plus (dashed line) versus exact simulations (solid line) for various spectral exponents $s$ of a sub-ohmic spin boson model. Parameters are: $\epsilon = 0$, $\Delta = 1$, $\omega_c = 20$, $T = 0$, and $\alpha = 0.05$.}
\label{f2}
\end{figure}

For a sub-ohmic bosonic bath, the mode distribution characterized by the parameter $s$ plays a significant role in the feasibility of a perturbative expansion according to Redfield-plus, as we show here. By fixing $\alpha = 0.05$,  in Fig.~\ref{f2}, we observe that the Redfield-plus can nearly reproduce the exact dynamics only for exponents $0.5\lesssim s$. However, as the weight of low-frequency modes increases with lowering $s$, the second order approximation becomes inadequate already on gradually shorter time scales.

\subsection{Time-depenmdemnt Memory Kernel for Populations: NIBA versus FP-HEOM}

For a spin system with no bias, i.e.\  $\epsilon=0$, immersed in a bosonic bath, the expectation is that the steady-state population is uniformly distributed between the two spin states, i.e., $P(t\to \infty) = \langle \sigma_z(t\to \infty) \rangle \rightarrow 0$. However, it is well-known that at zero temperature a symmetry breaking can take place corresponding to a quantum phase transition from a delocalized to localized asymptotic state \cite{kast2013persistence,xu2022taming}. Consequently, depending on the initial condition and reservoir parameters, $P(t) \neq 0$ over long timescales, which is a purely quantum phenomenon due to the absence of thermal fluctuations at zero temperature.
\begin{figure}[!htbp]
\centering
\includegraphics[width=8.5cm]{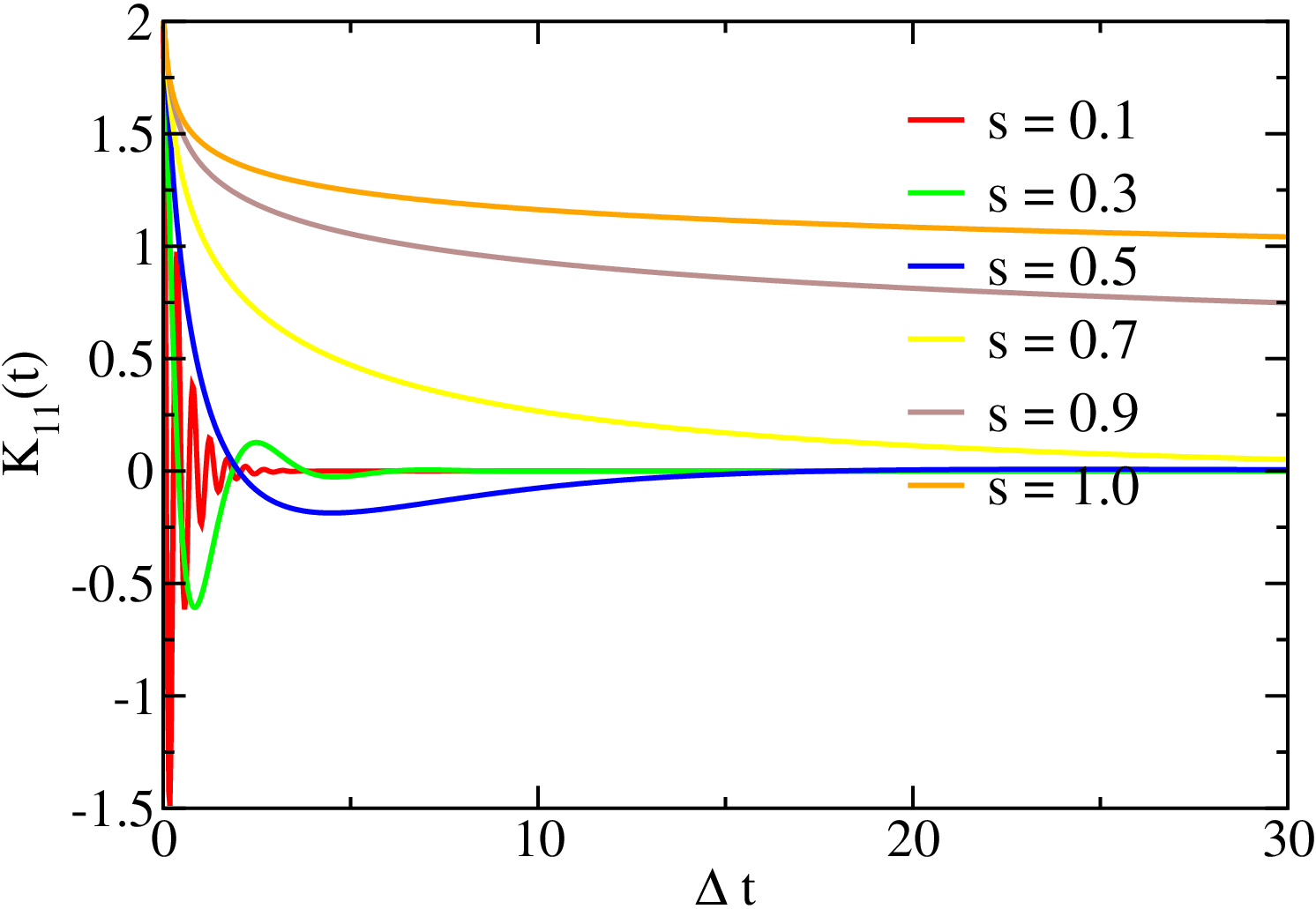}
\caption{The memory kernel in NIBA for various $s$. The simulation parameters are: $\epsilon = 0$, $\Delta = 1$, $\omega_c = 20$, $T = 0$, $\alpha = 0.05$.}
\label{f3}
\end{figure}

This behavior can be conveniently analyzed using the time-dependent memory kernel $K(t)$ which determines the spin dynamics according to
\begin{equation}\label{Eq:gme}
   \dot{P}_{\sigma}(t) = -\sum_{\sigma' = \pm}\int_0^t d\tau K_{\sigma,\sigma'}(t-\tau) P_{\sigma'}(\tau) \;\;.
\end{equation}

We mention here that a powerful perturbative treatment to derive the kernel is the so-called Non-interacting Blip Approximation (NIBA) \cite{dekker86,weiss12} and its extensions. There, kernels in powers of the tunnel splitting $\Delta$ are obtained with the NIBA kernel being of second order in the tunnel splitting $\Delta^2$. Accordingly, the NIBA is not based on a series expansion in $\alpha$ and thus accounts also for strong spin-bath coupling. It neglects long-range quantum coherences though and does not predict the correct equilibrium state for $\epsilon\neq 0$. 
In Fig.~\ref{f3} results for the NIBA memory kernel are depicted. From these data one would conclude a change in the dynamical behavior (monotonous decay versus oscillatory decay) to occur for values around $s=0.5$. For smaller exponents strong oscillations emerge with decreasing $s$.
\begin{figure}[!htbp]
\centering
\includegraphics[width=8.5cm]{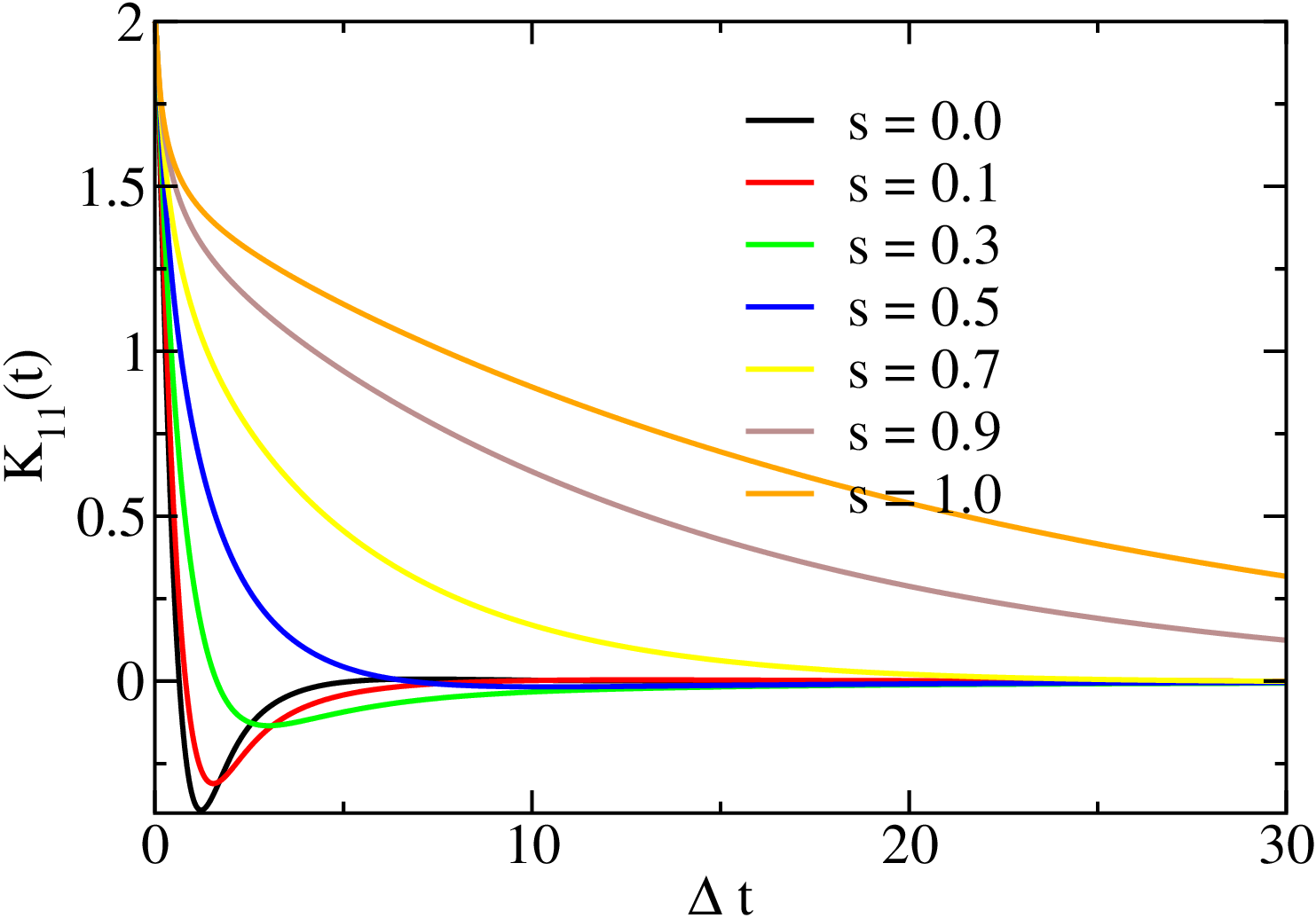}
\caption{The exact memory kernel in master equation in various $s$. The simulation parameters are: $\epsilon = 0$, $\Delta = 1$, $\omega_c = 20$, $T = 0$, $\alpha = 0.05$.}
\label{f4}
\end{figure}

In contrast, exact FP-HEOM data extracted from the exact dynamics \cite{xu2018convergence} are shown in Fig.~\ref{f4}. For the chosen coupling strength, it is also observed that for $s>0.5$, the memory decays clearly monotonous over time and remains always positive as also predicted by NIBA. However, quantitatively the exact decay appears to be much faster than within NIBA. For values of $s$ below this threshold oscillatory pattern are seen as well for the FP-HEOM results, but with much smoother and with less oscillations as the NIBA prediction. We can thus conclude that while the NIBA provides qualitatively the correct physics, it quantitatively in this regime of paramneter space not reliable.

Physically, the changeover in dynamical behavior in $K(t)$ can be attributed to a freezing of the population since the total weight of the kernel tends towards zero when integrated over a time span where $P_\sigma(t)$ does not change considerably. More specifically, in this regime of sufficiently small $s$, one may write for long times with $P_\sigma(\tau)\approx P_\sigma(t)$ in Eq. (\ref{Eq:gme}) that
\begin{equation}
   \lim_{t\to \infty} \dot{P}_{\sigma}(t) \approx -\sum_{\sigma' = \pm} k_{\sigma,\sigma'} P_{\sigma'}(t) \;\;.
\end{equation}
with rates
\begin{equation}
    k_{\sigma, \sigma'}= \int_0^\infty d\tau \ K_{\sigma,\sigma'}(\tau)\, .
\end{equation}

Estimations from Fig.~\ref{f4} allow to see that rates $k_{\sigma,\sigma'}$ tend indeed to zero corresponding to a slowing down of the relaxation dynamics up to the regime, where $k_{\sigma, \sigma'}=0$ so that the spin requires an infinitely long time to reach a Gibbs equilibrium state and instead displays, localization. This property aligns with the fact that hybridization between spin and reservoir induced by slow modes occurs, thus freezing the spin dynamics.

\section{Conclusion}

Since conventional second-order master equations fail in accurately capturing non-Markovian dynamics, this limitation becomes markedly significant for unconventional baths, particularly in conditions such as zero-temperature structured environments. One approach to circumvent this is to incorporate high-order corrections. However, the numerical computation of high-dimensional integrals in time poses considerable challenge as the accuracy of calculations become extremely sensitive to numerical errors. 

In this work, we have employed the Hierarchical Equations of Motion (HEOM) approach to effectively address these challenges. This methodology enables the systematic unraveling of higher-order master equations, balancing numerical efficiency with high precision. Consequently, our investigation has concentrated on the zero-temperature sub-Ohmic regimes, providing relevant computations to underscore our approach.

Furthermore, it is significant to note that the exact memory kernels in the quantum master equation can be extracted from the FP-HEOM approach and then be compared with those from perturbative schemes such as the NIBA. The extraction further  offers a useful tool for the analysis of quantum phase transition dynamics, establishing a comprehensive and reliable platform for further explorations into open quantum system dynamics.

\section*{Acknowledgment}
We thank Jiajun Ren, Haobin Wang, Frithjof B. Anders, and Matthias Vojta for sharing us their data as well as the fruitful discussions. M. X. acknowledges support from the state of BadenWürttemberg through bwHPC (JUSTUS 2). This work has been supported by the IQST, the German Science Foundation (DFG) under AN336/12-1 (For2724), the State of Baden-Wüttemberg under KQCBW/SiQuRe, the BadenWürttemberg Foundation within QT.BW (CDINQUA), and the BMBF through QSolid. 

\section*{Data availability}
The data that support the figures within this article are available from the corresponding author upon reasonable request

\bibliography{sn-bibliography}


\begin{thebibliography}{63}
\ifx \bisbn   \undefined \def \bisbn  #1{ISBN #1}\fi
\ifx \binits  \undefined \def \binits#1{#1}\fi
\ifx \bauthor  \undefined \def \bauthor#1{#1}\fi
\ifx \batitle  \undefined \def \batitle#1{#1}\fi
\ifx \bjtitle  \undefined \def \bjtitle#1{#1}\fi
\ifx \bvolume  \undefined \def \bvolume#1{\textbf{#1}}\fi
\ifx \byear  \undefined \def \byear#1{#1}\fi
\ifx \bissue  \undefined \def \bissue#1{#1}\fi
\ifx \bfpage  \undefined \def \bfpage#1{#1}\fi
\ifx \blpage  \undefined \def \blpage #1{#1}\fi
\ifx \burl  \undefined \def \burl#1{\textsf{#1}}\fi
\ifx \doiurl  \undefined \def \doiurl#1{\url{https://doi.org/#1}}\fi
\ifx \betal  \undefined \def \betal{\textit{et al.}}\fi
\ifx \binstitute  \undefined \def \binstitute#1{#1}\fi
\ifx \binstitutionaled  \undefined \def \binstitutionaled#1{#1}\fi
\ifx \bctitle  \undefined \def \bctitle#1{#1}\fi
\ifx \beditor  \undefined \def \beditor#1{#1}\fi
\ifx \bpublisher  \undefined \def \bpublisher#1{#1}\fi
\ifx \bbtitle  \undefined \def \bbtitle#1{#1}\fi
\ifx \bedition  \undefined \def \bedition#1{#1}\fi
\ifx \bseriesno  \undefined \def \bseriesno#1{#1}\fi
\ifx \blocation  \undefined \def \blocation#1{#1}\fi
\ifx \bsertitle  \undefined \def \bsertitle#1{#1}\fi
\ifx \bsnm \undefined \def \bsnm#1{#1}\fi
\ifx \bsuffix \undefined \def \bsuffix#1{#1}\fi
\ifx \bparticle \undefined \def \bparticle#1{#1}\fi
\ifx \barticle \undefined \def \barticle#1{#1}\fi
\bibcommenthead
\ifx \bconfdate \undefined \def \bconfdate #1{#1}\fi
\ifx \botherref \undefined \def \botherref #1{#1}\fi
\ifx \url \undefined \def \url#1{\textsf{#1}}\fi
\ifx \bchapter \undefined \def \bchapter#1{#1}\fi
\ifx \bbook \undefined \def \bbook#1{#1}\fi
\ifx \bcomment \undefined \def \bcomment#1{#1}\fi
\ifx \oauthor \undefined \def \oauthor#1{#1}\fi
\ifx \citeauthoryear \undefined \def \citeauthoryear#1{#1}\fi
\ifx \endbibitem  \undefined \def \endbibitem {}\fi
\ifx \bconflocation  \undefined \def \bconflocation#1{#1}\fi
\ifx \arxivurl  \undefined \def \arxivurl#1{\textsf{#1}}\fi
\csname PreBibitemsHook\endcsname

\bibitem[\protect\citeauthoryear{Gardiner and
  Zoller}{2004}]{gardiner2004quantum}
\begin{bbook}
\bauthor{\bsnm{Gardiner}, \binits{C.W.}},
\bauthor{\bsnm{Zoller}, \binits{P.}}:
\bbtitle{Quantum Noise: a Handbook of Markovian and non-Markovian Quantum
  Stochastic Methods with Applications to Quantum Optics},
\bedition{3rd} edn.
\bpublisher{Springer},
\blocation{Berlin}
(\byear{2004})
\end{bbook}
\endbibitem

\bibitem[\protect\citeauthoryear{Scully and Zubairy}{1997}]{scully1997quantum}
\begin{bbook}
\bauthor{\bsnm{Scully}, \binits{M.O.}},
\bauthor{\bsnm{Zubairy}, \binits{M.S.}}:
\bbtitle{Quantum Optics}.
\bpublisher{Cambridge University Press}, \blocation{???}
(\byear{1997}).
\doiurl{10.1017/CBO9780511813993}
\end{bbook}
\endbibitem

\bibitem[\protect\citeauthoryear{Kamenev}{2023}]{kamenev2023field}
\begin{bbook}
\bauthor{\bsnm{Kamenev}, \binits{A.}}:
\bbtitle{Field Theory of Non-equilibrium Systems}.
\bpublisher{Cambridge University Press}, \blocation{???}
(\byear{2023})
\end{bbook}
\endbibitem

\bibitem[\protect\citeauthoryear{Nitzan}{2006}]{nitzan06}
\begin{bbook}
\bauthor{\bsnm{Nitzan}, \binits{A.}}:
\bbtitle{Chemical Dynamics in Condensed Phases}.
\bpublisher{Oxford University Press},
\blocation{New York}
(\byear{2006})
\end{bbook}
\endbibitem

\bibitem[\protect\citeauthoryear{May and K\"{u}hn}{2011}]{may11}
\begin{bbook}
\bauthor{\bsnm{May}, \binits{V.}},
\bauthor{\bsnm{K\"{u}hn}, \binits{O.}}:
\bbtitle{Charge and Energy Transfer Dynamics in Molecular Systems},
\bedition{3}rd edn.
\bpublisher{Wiley-VCH},
\blocation{Weinheim}
(\byear{2011})
\end{bbook}
\endbibitem

\bibitem[\protect\citeauthoryear{Mohseni et~al.}{2014}]{mohseni2014quantum}
\begin{bbook}
\bauthor{\bsnm{Mohseni}, \binits{M.}},
\bauthor{\bsnm{Omar}, \binits{Y.}},
\bauthor{\bsnm{Engel}, \binits{G.S.}},
\bauthor{\bsnm{Plenio}, \binits{M.B.}}:
\bbtitle{Quantum Effects in Biology}.
\bpublisher{Cambridge University Press}, \blocation{???}
(\byear{2014})
\end{bbook}
\endbibitem

\bibitem[\protect\citeauthoryear{Weiss}{2012}]{weiss12}
\begin{bbook}
\bauthor{\bsnm{Weiss}, \binits{U.}}:
\bbtitle{Quantum Dissipative Systems},
\bedition{4th} edn.
\bpublisher{World Scientific},
\blocation{New Jersey}
(\byear{2012})
\end{bbook}
\endbibitem

\bibitem[\protect\citeauthoryear{Kast and
  Ankerhold}{2013}]{kast2013persistence}
\begin{barticle}
\bauthor{\bsnm{Kast}, \binits{D.}},
\bauthor{\bsnm{Ankerhold}, \binits{J.}}:
\batitle{Persistence of coherent quantum dynamics at strong dissipation}.
\bjtitle{Phys. Rev. Lett.}
\bvolume{110},
\bfpage{010402}
(\byear{2013})
\doiurl{10.1103/PhysRevLett.110.010402}
\end{barticle}
\endbibitem

\bibitem[\protect\citeauthoryear{Xu et~al.}{2022}]{xu2022taming}
\begin{barticle}
\bauthor{\bsnm{Xu}, \binits{M.}},
\bauthor{\bsnm{Yan}, \binits{Y.}},
\bauthor{\bsnm{Shi}, \binits{Q.}},
\bauthor{\bsnm{Ankerhold}, \binits{J.}},
\bauthor{\bsnm{Stockburger}, \binits{J.T.}}:
\batitle{Taming quantum noise for efficient low temperature simulations of open
  quantum systems}.
\bjtitle{Phys. Rev. Lett.}
\bvolume{129},
\bfpage{230601}
(\byear{2022})
\doiurl{10.1103/PhysRevLett.129.230601}
\end{barticle}
\endbibitem

\bibitem[\protect\citeauthoryear{Zhang and Yan}{2016}]{zhang16}
\begin{barticle}
\bauthor{\bsnm{Zhang}, \binits{H.-D.}},
\bauthor{\bsnm{Yan}, \binits{Y.-J.}}:
\batitle{Kinetic rate kernels via hierarchical liouville-space projection
  operator approach}.
\bjtitle{J.~Phys.~Chem.~A}
\bvolume{120},
\bfpage{3241}--\blpage{3245}
(\byear{2016})
\end{barticle}
\endbibitem

\bibitem[\protect\citeauthoryear{Xu et~al.}{2017}]{xu17}
\begin{barticle}
\bauthor{\bsnm{Xu}, \binits{M.}},
\bauthor{\bsnm{Song}, \binits{L.}},
\bauthor{\bsnm{Song}, \binits{K.}},
\bauthor{\bsnm{Shi}, \binits{Q.}}:
\batitle{Convergence of high order perturbative expansions in open system
  quantum dynamics}.
\bjtitle{J.~Chem.~Phys.}
\bvolume{146}(\bissue{6}),
\bfpage{064102}
(\byear{2017})
\end{barticle}
\endbibitem

\bibitem[\protect\citeauthoryear{Xu et~al.}{2018}]{xu2018convergence}
\begin{barticle}
\bauthor{\bsnm{Xu}, \binits{M.}},
\bauthor{\bsnm{Yan}, \binits{Y.}},
\bauthor{\bsnm{Liu}, \binits{Y.}},
\bauthor{\bsnm{Shi}, \binits{Q.}}:
\batitle{Convergence of high order memory kernels in the nakajima-zwanzig
  generalized master equation and rate constants: Case study of the spin-boson
  model}.
\bjtitle{J.~Chem.~Phys.}
\bvolume{148}(\bissue{16}),
\bfpage{164101}
(\byear{2018})
\end{barticle}
\endbibitem

\bibitem[\protect\citeauthoryear{Shibata and Arimitsu}{1980}]{shibata80}
\begin{barticle}
\bauthor{\bsnm{Shibata}, \binits{F.}},
\bauthor{\bsnm{Arimitsu}, \binits{T.}}:
\batitle{Expansion formulas in nonequilibrium statistical mechanics}.
\bjtitle{J. Phys. Soc. Jpn.}
\bvolume{49}(\bissue{3}),
\bfpage{891}--\blpage{897}
(\byear{1980})
\end{barticle}
\endbibitem

\bibitem[\protect\citeauthoryear{Terwiel}{1974}]{terwiel74}
\begin{barticle}
\bauthor{\bsnm{Terwiel}, \binits{R.H.}}:
\batitle{Projection operator method applied to stochastic linear differential
  equations}.
\bjtitle{Physica}
\bvolume{74}(\bissue{2}),
\bfpage{248}--\blpage{265}
(\byear{1974})
\end{barticle}
\endbibitem

\bibitem[\protect\citeauthoryear{Gong et~al.}{2015}]{gong15}
\begin{barticle}
\bauthor{\bsnm{Gong}, \binits{Z.}},
\bauthor{\bsnm{Tang}, \binits{Z.}},
\bauthor{\bsnm{Mukamel}, \binits{S.}},
\bauthor{\bsnm{Cao}, \binits{J.}},
\bauthor{\bsnm{Wu}, \binits{J.}}:
\batitle{A continued function resummation form of bath relaxation effect in the
  spin-boson model}.
\bjtitle{J.~Chem.~Phys.}
\bvolume{142},
\bfpage{084103}
(\byear{2015})
\end{barticle}
\endbibitem

\bibitem[\protect\citeauthoryear{Aihara et~al.}{1990}]{aihara90}
\begin{barticle}
\bauthor{\bsnm{Aihara}, \binits{M.}},
\bauthor{\bsnm{Sevian}, \binits{H.M.}},
\bauthor{\bsnm{Skinner}, \binits{J.L.}}:
\batitle{Non-markovian relaxation of a spin-1/2 particle in a fluctuating
  transverse field: Cumulant expansion and stochastic simulation results}.
\bjtitle{Phys. Rev. A}
\bvolume{41},
\bfpage{6596}
(\byear{1990})
\end{barticle}
\endbibitem

\bibitem[\protect\citeauthoryear{Tanimura and Kubo}{1989}]{tanimura89}
\begin{barticle}
\bauthor{\bsnm{Tanimura}, \binits{Y.}},
\bauthor{\bsnm{Kubo}, \binits{R.}}:
\batitle{Time evolution of a quantum system in contact with a nearly
  gaussian-markoffian noise bath}.
\bjtitle{J. Phys. Soc. Jpn.}
\bvolume{58},
\bfpage{101}
(\byear{1989})
\end{barticle}
\endbibitem

\bibitem[\protect\citeauthoryear{Tanimura}{2006}]{tanimura06}
\begin{barticle}
\bauthor{\bsnm{Tanimura}, \binits{Y.}}:
\batitle{Stochastic liouville, langevin, {Fokker}-{Planck}, and master equation
  approaches to quantum dissipative systems}.
\bjtitle{J. Phys. Soc. Jpn.}
\bvolume{75},
\bfpage{082001}--\blpage{082039}
(\byear{2006})
\end{barticle}
\endbibitem

\bibitem[\protect\citeauthoryear{Tanimura}{2020}]{tanimura2020numerically}
\begin{barticle}
\bauthor{\bsnm{Tanimura}, \binits{Y.}}:
\batitle{Numerically “exact” approach to open quantum dynamics: The
  hierarchical equations of motion (heom)}.
\bjtitle{J.~Chem.~Phys.}
\bvolume{153}(\bissue{2}),
\bfpage{020901}
(\byear{2020})
\end{barticle}
\endbibitem

\bibitem[\protect\citeauthoryear{Zhang et~al.}{2020}]{zhang2020hierarchical}
\begin{barticle}
\bauthor{\bsnm{Zhang}, \binits{H.-D.}},
\bauthor{\bsnm{Cui}, \binits{L.}},
\bauthor{\bsnm{Gong}, \binits{H.}},
\bauthor{\bsnm{Xu}, \binits{R.-X.}},
\bauthor{\bsnm{Zheng}, \binits{X.}},
\bauthor{\bsnm{Yan}, \binits{Y.}}:
\batitle{Hierarchical equations of motion method based on fano spectrum
  decomposition for low temperature environments}.
\bjtitle{J.~Chem.~Phys.}
\bvolume{152}(\bissue{6}),
\bfpage{064107}
(\byear{2020})
\end{barticle}
\endbibitem

\bibitem[\protect\citeauthoryear{Hsieh and Cao}{2018}]{hsieh2018unified}
\begin{barticle}
\bauthor{\bsnm{Hsieh}, \binits{C.-Y.}},
\bauthor{\bsnm{Cao}, \binits{J.}}:
\batitle{A unified stochastic formulation of dissipative quantum dynamics. i.
  generalized hierarchical equations}.
\bjtitle{J.~Chem.~Phys.}
\bvolume{148}(\bissue{1}),
\bfpage{014103}
(\byear{2018})
\end{barticle}
\endbibitem

\bibitem[\protect\citeauthoryear{Wang et~al.}{2019}]{wang2019dynamical}
\begin{barticle}
\bauthor{\bsnm{Wang}, \binits{Q.}},
\bauthor{\bsnm{Gong}, \binits{Z.}},
\bauthor{\bsnm{Duan}, \binits{C.}},
\bauthor{\bsnm{Tang}, \binits{Z.}},
\bauthor{\bsnm{Wu}, \binits{J.}}:
\batitle{Dynamical scaling in the ohmic spin-boson model studied by extended
  hierarchical equations of motion}.
\bjtitle{J.~Chem.~Phys.}
\bvolume{150}(\bissue{8}),
\bfpage{084114}
(\byear{2019})
\end{barticle}
\endbibitem

\bibitem[\protect\citeauthoryear{Shi et~al.}{2009}]{shi09b}
\begin{barticle}
\bauthor{\bsnm{Shi}, \binits{Q.}},
\bauthor{\bsnm{Chen}, \binits{L.-P.}},
\bauthor{\bsnm{Nan}, \binits{G.-J.}},
\bauthor{\bsnm{Xu}, \binits{R.-X.}},
\bauthor{\bsnm{Yan}, \binits{Y.-J.}}:
\batitle{Efficient hierarchical liouville--space propagator to quantum
  dissipative dynamics}.
\bjtitle{J. Chem. Phys.}
\bvolume{130},
\bfpage{084105}--\blpage{084108}
(\byear{2009})
\end{barticle}
\endbibitem

\bibitem[\protect\citeauthoryear{Liu et~al.}{2014}]{liu14}
\begin{barticle}
\bauthor{\bsnm{Liu}, \binits{H.}},
\bauthor{\bsnm{Zhu}, \binits{L.}},
\bauthor{\bsnm{Bai}, \binits{S.}},
\bauthor{\bsnm{Shi}, \binits{Q.}}:
\batitle{{Reduced Quantum Dynamics with Arbitrary Bath Spectral Densities:
  Hierarchical Equations of Motion Based on Several Different Bath
  Decomposition Schemes.}}
\bjtitle{J. Chem. Phys.}
\bvolume{140},
\bfpage{134106}
(\byear{2014})
\end{barticle}
\endbibitem

\bibitem[\protect\citeauthoryear{Shi et~al.}{2018}]{shi2018efficient}
\begin{barticle}
\bauthor{\bsnm{Shi}, \binits{Q.}},
\bauthor{\bsnm{Xu}, \binits{Y.}},
\bauthor{\bsnm{Yan}, \binits{Y.}},
\bauthor{\bsnm{Xu}, \binits{M.}}:
\batitle{Efficient propagation of the hierarchical equations of motion using
  the matrix product state method}.
\bjtitle{J.~Chem.~Phys.}
\bvolume{148}(\bissue{17}),
\bfpage{174102}
(\byear{2018})
\end{barticle}
\endbibitem

\bibitem[\protect\citeauthoryear{Han et~al.}{2019}]{han2019stochastic}
\begin{barticle}
\bauthor{\bsnm{Han}, \binits{L.}},
\bauthor{\bsnm{Chernyak}, \binits{V.}},
\bauthor{\bsnm{Yan}, \binits{Y.-A.}},
\bauthor{\bsnm{Zheng}, \binits{X.}},
\bauthor{\bsnm{Yan}, \binits{Y.}}:
\batitle{Stochastic representation of non-markovian fermionic quantum
  dissipation}.
\bjtitle{Phys. Rev. Lett.}
\bvolume{123},
\bfpage{050601}
(\byear{2019})
\doiurl{10.1103/PhysRevLett.123.050601}
\end{barticle}
\endbibitem

\bibitem[\protect\citeauthoryear{Yan et~al.}{2019}]{yan2019unified}
\begin{barticle}
\bauthor{\bsnm{Yan}, \binits{Y.-A.}},
\bauthor{\bsnm{Wang}, \binits{H.}},
\bauthor{\bsnm{Shao}, \binits{J.}}:
\batitle{A unified view of hierarchy approach and formula of differentiation}.
\bjtitle{J.~Chem.~Phys.}
\bvolume{151}(\bissue{16}),
\bfpage{164110}
(\byear{2019})
\end{barticle}
\endbibitem

\bibitem[\protect\citeauthoryear{Erpenbeck
  et~al.}{2018}]{erpenbeck2018extending}
\begin{barticle}
\bauthor{\bsnm{Erpenbeck}, \binits{A.}},
\bauthor{\bsnm{Hertlein}, \binits{C.}},
\bauthor{\bsnm{Schinabeck}, \binits{C.}},
\bauthor{\bsnm{Thoss}, \binits{M.}}:
\batitle{Extending the hierarchical quantum master equation approach to low
  temperatures and realistic band structures}.
\bjtitle{J.~Chem.~Phys.}
\bvolume{149}(\bissue{6}),
\bfpage{064106}
(\byear{2018})
\end{barticle}
\endbibitem

\bibitem[\protect\citeauthoryear{Rahman and
  Kleinekath{\"o}fer}{2019}]{rahman2019chebyshev}
\begin{barticle}
\bauthor{\bsnm{Rahman}, \binits{H.}},
\bauthor{\bsnm{Kleinekath{\"o}fer}, \binits{U.}}:
\batitle{Chebyshev hierarchical equations of motion for systems with arbitrary
  spectral densities and temperatures}.
\bjtitle{J.~Chem.~Phys.}
\bvolume{150}(\bissue{24}),
\bfpage{244104}
(\byear{2019})
\end{barticle}
\endbibitem

\bibitem[\protect\citeauthoryear{Yan et~al.}{2020}]{yan2020new}
\begin{barticle}
\bauthor{\bsnm{Yan}, \binits{Y.}},
\bauthor{\bsnm{Xing}, \binits{T.}},
\bauthor{\bsnm{Shi}, \binits{Q.}}:
\batitle{A new method to improve the numerical stability of the hierarchical
  equations of motion for discrete harmonic oscillator modes}.
\bjtitle{J.~Chem.~Phys.}
\bvolume{153}(\bissue{20}),
\bfpage{204109}
(\byear{2020})
\end{barticle}
\endbibitem

\bibitem[\protect\citeauthoryear{Ikeda and
  Scholes}{2020}]{ikeda2020generalization}
\begin{barticle}
\bauthor{\bsnm{Ikeda}, \binits{T.}},
\bauthor{\bsnm{Scholes}, \binits{G.D.}}:
\batitle{Generalization of the hierarchical equations of motion theory for
  efficient calculations with arbitrary correlation functions}.
\bjtitle{J.~Chem.~Phys.}
\bvolume{152}(\bissue{20}),
\bfpage{204101}
(\byear{2020})
\end{barticle}
\endbibitem

\bibitem[\protect\citeauthoryear{Nakamura and Tanimura}{2018}]{nakamura18}
\begin{barticle}
\bauthor{\bsnm{Nakamura}, \binits{K.}},
\bauthor{\bsnm{Tanimura}, \binits{Y.}}:
\batitle{Hierarchical schr\"odinger equations of motion for open quantum
  dynamics}.
\bjtitle{Phys. Rev. A}
\bvolume{98},
\bfpage{012109}
(\byear{2018})
\doiurl{10.1103/PhysRevA.98.012109}
\end{barticle}
\endbibitem

\bibitem[\protect\citeauthoryear{H\"{a}rtle et~al.}{2013}]{hartle13}
\begin{barticle}
\bauthor{\bsnm{H\"{a}rtle}, \binits{R.}},
\bauthor{\bsnm{Cohen}, \binits{G.}},
\bauthor{\bsnm{Reichman}, \binits{D.R.}},
\bauthor{\bsnm{Millis}, \binits{A.J.}}:
\batitle{Decoherence and lead-induced interdot coupling in nonequilibrium
  electron transport through interacting quantum dots: A hierarchical quantum
  master equation approach}.
\bjtitle{Phys.~Rev.~B}
\bvolume{88},
\bfpage{235426}
(\byear{2013})
\end{barticle}
\endbibitem

\bibitem[\protect\citeauthoryear{Dunn et~al.}{2019}]{dunn2019removing}
\begin{barticle}
\bauthor{\bsnm{Dunn}, \binits{I.S.}},
\bauthor{\bsnm{Tempelaar}, \binits{R.}},
\bauthor{\bsnm{Reichman}, \binits{D.R.}}:
\batitle{Removing instabilities in the hierarchical equations of motion: Exact
  and approximate projection approaches}.
\bjtitle{J.~Chem.~Phys.}
\bvolume{150}(\bissue{18}),
\bfpage{184109}
(\byear{2019})
\end{barticle}
\endbibitem

\bibitem[\protect\citeauthoryear{Yan and Shao}{2016}]{yan2016stochastic}
\begin{barticle}
\bauthor{\bsnm{Yan}, \binits{Y.-A.}},
\bauthor{\bsnm{Shao}, \binits{J.}}:
\batitle{Stochastic description of quantum brownian dynamics}.
\bjtitle{Frontiers of Physics}
\bvolume{11}(\bissue{4}),
\bfpage{1}--\blpage{24}
(\byear{2016})
\end{barticle}
\endbibitem

\bibitem[\protect\citeauthoryear{Ke et~al.}{2022}]{ke2022hierarchical}
\begin{barticle}
\bauthor{\bsnm{Ke}, \binits{Y.}},
\bauthor{\bsnm{Borrelli}, \binits{R.}},
\bauthor{\bsnm{Thoss}, \binits{M.}}:
\batitle{Hierarchical equations of motion approach to hybrid fermionic and
  bosonic environments: Matrix product state formulation in twin space}.
\bjtitle{J.~Chem.~Phys.}
\bvolume{156}(\bissue{19}),
\bfpage{194102}
(\byear{2022})
\end{barticle}
\endbibitem

\bibitem[\protect\citeauthoryear{Li et~al.}{2023}]{li2023dissipatons}
\begin{botherref}
\oauthor{\bsnm{Li}, \binits{X.}},
\oauthor{\bsnm{Su}, \binits{Y.}},
\oauthor{\bsnm{Chen}, \binits{Z.-H.}},
\oauthor{\bsnm{Wang}, \binits{Y.}},
\oauthor{\bsnm{Xu}, \binits{R.-X.}},
\oauthor{\bsnm{Zheng}, \binits{X.}},
\oauthor{\bsnm{Yan}, \binits{Y.}}:
Dissipatons as generalized brownian particles for open quantum systems:
  Dissipaton-embedded quantum master equation.
arXiv preprint arXiv:2303.10666
(2023)
\end{botherref}
\endbibitem

\bibitem[\protect\citeauthoryear{Yan et~al.}{2021}]{yan2021efficient}
\begin{barticle}
\bauthor{\bsnm{Yan}, \binits{Y.}},
\bauthor{\bsnm{Xu}, \binits{M.}},
\bauthor{\bsnm{Li}, \binits{T.}},
\bauthor{\bsnm{Shi}, \binits{Q.}}:
\batitle{Efficient propagation of the hierarchical equations of motion using
  the tucker and hierarchical tucker tensors}.
\bjtitle{J.~Chem.~Phys.}
\bvolume{154}(\bissue{19}),
\bfpage{194104}
(\byear{2021})
\end{barticle}
\endbibitem

\bibitem[\protect\citeauthoryear{Li et~al.}{2022}]{li2022low}
\begin{botherref}
\oauthor{\bsnm{Li}, \binits{T.}},
\oauthor{\bsnm{Yan}, \binits{Y.}},
\oauthor{\bsnm{Shi}, \binits{Q.}}:
A low-temperature quantum {Fokker}-{Planck} equation that improvesthe numerical
  stability of the hierarchical equations of motion for the brownian oscillator
  spectral density.
The Journal of Chemical Physics
(2022)
\end{botherref}
\endbibitem

\bibitem[\protect\citeauthoryear{Ke}{2023}]{ke2023tree}
\begin{botherref}
\oauthor{\bsnm{Ke}, \binits{Y.}}:
Tree tensor network state approach for solving hierarchical equations of
  motions.
arXiv preprint arXiv:2304.05151
(2023)
\end{botherref}
\endbibitem

\bibitem[\protect\citeauthoryear{Chen et~al.}{2022}]{chen2022universal}
\begin{botherref}
\oauthor{\bsnm{Chen}, \binits{Z.-H.}},
\oauthor{\bsnm{Wang}, \binits{Y.}},
\oauthor{\bsnm{Zheng}, \binits{X.}},
\oauthor{\bsnm{Xu}, \binits{R.-X.}},
\oauthor{\bsnm{Yan}, \binits{Y.}}:
Universal time-domain prony fitting decomposition for optimized hierarchical
  quantum master equations.
J.~Chem.~Phys.
(2022)
\end{botherref}
\endbibitem

\bibitem[\protect\citeauthoryear{Fay}{2022}]{fay2022simple}
\begin{barticle}
\bauthor{\bsnm{Fay}, \binits{T.P.}}:
\batitle{A simple improved low temperature correction for the hierarchical
  equations of motion}.
\bjtitle{The Journal of Chemical Physics}
\bvolume{157}(\bissue{5}),
\bfpage{054108}
(\byear{2022})
\end{barticle}
\endbibitem

\bibitem[\protect\citeauthoryear{Yan}{2014}]{yan14a}
\begin{barticle}
\bauthor{\bsnm{Yan}, \binits{Y.-J.}}:
\batitle{Theory of open quantum systems with bath of electrons and phonons and
  spins: Many-dissipaton density matrixes approach}.
\bjtitle{J.~Chem.~Phys.}
\bvolume{140},
\bfpage{054105}
(\byear{2014})
\end{barticle}
\endbibitem

\bibitem[\protect\citeauthoryear{Dan et~al.}{2022}]{dan2022efficient}
\begin{botherref}
\oauthor{\bsnm{Dan}, \binits{X.}},
\oauthor{\bsnm{Xu}, \binits{M.}},
\oauthor{\bsnm{Stockburger}, \binits{J.}},
\oauthor{\bsnm{Ankerhold}, \binits{J.}},
\oauthor{\bsnm{Shi}, \binits{Q.}}:
Efficient low temperature simulations for fermionic reservoirs with the
  hierarchical equations of motion method: Application to the anderson impurity
  model.
arXiv preprint arXiv:2211.04089
(2022)
\end{botherref}
\endbibitem

\bibitem[\protect\citeauthoryear{Tamascelli
  et~al.}{2018}]{tamascelli2018nonperturbative}
\begin{barticle}
\bauthor{\bsnm{Tamascelli}, \binits{D.}},
\bauthor{\bsnm{Smirne}, \binits{A.}},
\bauthor{\bsnm{Huelga}, \binits{S.F.}},
\bauthor{\bsnm{Plenio}, \binits{M.B.}}:
\batitle{Nonperturbative treatment of non-markovian dynamics of open quantum
  systems}.
\bjtitle{Phys.~Rev.~Lett.}
\bvolume{120}(\bissue{3}),
\bfpage{030402}
(\byear{2018})
\end{barticle}
\endbibitem

\bibitem[\protect\citeauthoryear{Suess et~al.}{2014}]{suess14}
\begin{barticle}
\bauthor{\bsnm{Suess}, \binits{D.}},
\bauthor{\bsnm{Eisfeld}, \binits{A.}},
\bauthor{\bsnm{Strunz}, \binits{W.T.}}:
\batitle{Hierarchy of stochastic pure states for open quantum system dynamics}.
\bjtitle{Phys.~Rev.~Lett.}
\bvolume{113}(\bissue{15}),
(\byear{2014})
\end{barticle}
\endbibitem

\bibitem[\protect\citeauthoryear{Arrigoni
  et~al.}{2013}]{arrigoni2013nonequilibrium}
\begin{barticle}
\bauthor{\bsnm{Arrigoni}, \binits{E.}},
\bauthor{\bsnm{Knap}, \binits{M.}},
\bauthor{\bsnm{Von Der~Linden}, \binits{W.}}:
\batitle{Nonequilibrium dynamical mean-field theory: An auxiliary quantum
  master equation approach}.
\bjtitle{Phys.~Rev.~Lett.}
\bvolume{110}(\bissue{8}),
\bfpage{086403}
(\byear{2013})
\end{barticle}
\endbibitem

\bibitem[\protect\citeauthoryear{Leggett et~al.}{1987}]{leggett87}
\begin{barticle}
\bauthor{\bsnm{Leggett}, \binits{A.J.}},
\bauthor{\bsnm{Chakravarty}, \binits{S.}},
\bauthor{\bsnm{Dorsey}, \binits{A.T.}},
\bauthor{\bsnm{Fisher}, \binits{M.P.A.}},
\bauthor{\bsnm{Garg}, \binits{A.}},
\bauthor{\bsnm{Zwerger}, \binits{W.}}:
\batitle{Dynamics of the dissipative two-level system}.
\bjtitle{Rev.~Mod.~Phys.}
\bvolume{59},
\bfpage{1}
(\byear{1987})
\end{barticle}
\endbibitem

\bibitem[\protect\citeauthoryear{Tanimura}{2014}]{tanimura14}
\begin{barticle}
\bauthor{\bsnm{Tanimura}, \binits{Y.}}:
\batitle{Reduced hierarchical equations of motion in real and imaginary time:
  Correlated initial states and thermodynamic quantities}.
\bjtitle{J.~Chem.~Phys.}
\bvolume{141},
\bfpage{044114}
(\byear{2014})
\end{barticle}
\endbibitem

\bibitem[\protect\citeauthoryear{Song and Shi}{2015}]{song15b}
\begin{barticle}
\bauthor{\bsnm{Song}, \binits{L.Z.}},
\bauthor{\bsnm{Shi}, \binits{Q.}}:
\batitle{Calculation of correlated initial state in the hierarchical equations
  of motion method using an imaginary time path integral approach}.
\bjtitle{J.~Chem.~Phys.}
\bvolume{143},
\bfpage{194106}
(\byear{2015})
\end{barticle}
\endbibitem

\bibitem[\protect\citeauthoryear{Feynman and Vernon}{1963}]{feynman63}
\begin{barticle}
\bauthor{\bsnm{Feynman}, \binits{R.P.}},
\bauthor{\bsnm{Vernon}, \binits{F.L.}}:
\batitle{The theory of a general quantum system interacting with a linear
  dissipative system}.
\bjtitle{Ann. Phys.}
\bvolume{24},
\bfpage{118}
(\byear{1963})
\end{barticle}
\endbibitem

\bibitem[\protect\citeauthoryear{M{\"u}hlbacher
  et~al.}{2004}]{muhlbacher2004path}
\begin{barticle}
\bauthor{\bsnm{M{\"u}hlbacher}, \binits{L.}},
\bauthor{\bsnm{Ankerhold}, \binits{J.}},
\bauthor{\bsnm{Escher}, \binits{C.}}:
\batitle{Path-integral monte carlo simulations for electronic dynamics on
  molecular chains. i. sequential hopping and super exchange}.
\bjtitle{J.~Chem.~Phys.}
\bvolume{121}(\bissue{24}),
\bfpage{12696}--\blpage{12707}
(\byear{2004})
\end{barticle}
\endbibitem

\bibitem[\protect\citeauthoryear{Nakatsukasa et~al.}{2018}]{nakatsukasa2018aaa}
\begin{barticle}
\bauthor{\bsnm{Nakatsukasa}, \binits{Y.}},
\bauthor{\bsnm{S{\`e}te}, \binits{O.}},
\bauthor{\bsnm{Trefethen}, \binits{L.N.}}:
\batitle{The aaa algorithm for rational approximation}.
\bjtitle{SIAM J. Sci. Comput.}
\bvolume{40}(\bissue{3}),
\bfpage{1494}--\blpage{1522}
(\byear{2018})
\end{barticle}
\endbibitem

\bibitem[\protect\citeauthoryear{Xu et~al.}{2005}]{xu05}
\begin{barticle}
\bauthor{\bsnm{Xu}, \binits{R.-X.}},
\bauthor{\bsnm{Cui}, \binits{P.}},
\bauthor{\bsnm{Li}, \binits{X.-Q.}},
\bauthor{\bsnm{Mo}, \binits{Y.}},
\bauthor{\bsnm{Yan}, \binits{Y.-J.}}:
\batitle{Exact quantum master equation via the calculus on path integrals}.
\bjtitle{J.~Chem.~Phys.}
\bvolume{122},
\bfpage{041103}
(\byear{2005})
\end{barticle}
\endbibitem

\bibitem[\protect\citeauthoryear{Xu et~al.}{2021}]{xu2021heat}
\begin{barticle}
\bauthor{\bsnm{Xu}, \binits{M.}},
\bauthor{\bsnm{Stockburger}, \binits{J.T.}},
\bauthor{\bsnm{Ankerhold}, \binits{J.}}:
\batitle{Heat transport through a superconducting artificial atom}.
\bjtitle{Phys. Rev. B}
\bvolume{103},
\bfpage{104304}
(\byear{2021})
\doiurl{10.1103/PhysRevB.103.104304}
\end{barticle}
\endbibitem

\bibitem[\protect\citeauthoryear{Xu et~al.}{2022}]{xu2022minimal}
\begin{barticle}
\bauthor{\bsnm{Xu}, \binits{M.}},
\bauthor{\bsnm{Stockburger}, \binits{J.}},
\bauthor{\bsnm{Kurizki}, \binits{G.}},
\bauthor{\bsnm{Ankerhold}, \binits{J.}}:
\batitle{Minimal quantum thermal machine in a bandgap environment:
  non-markovian features and anti-zeno advantage}.
\bjtitle{New.~J.~Phys.}
\bvolume{24}(\bissue{3}),
\bfpage{035003}
(\byear{2022})
\end{barticle}
\endbibitem

\bibitem[\protect\citeauthoryear{Lubich et~al.}{2015}]{lubich15}
\begin{barticle}
\bauthor{\bsnm{Lubich}, \binits{C.}},
\bauthor{\bsnm{Oseledets}, \binits{I.}},
\bauthor{\bsnm{Vandereycken}, \binits{B.}}:
\batitle{Time integration of tensor trains}.
\bjtitle{SIAM J. Num. Anal.}
\bvolume{53},
\bfpage{917}--\blpage{941}
(\byear{2015})
\end{barticle}
\endbibitem

\bibitem[\protect\citeauthoryear{Breuer and Petruccione}{2002}]{breuer02}
\begin{bbook}
\bauthor{\bsnm{Breuer}, \binits{H.P.}},
\bauthor{\bsnm{Petruccione}, \binits{F.}}:
\bbtitle{The Theory of Open Quantum Systems}.
\bpublisher{Oxford University Press},
\blocation{New York}
(\byear{2002})
\end{bbook}
\endbibitem

\bibitem[\protect\citeauthoryear{Traversa
  et~al.}{2013}]{traversa2013generalized}
\begin{barticle}
\bauthor{\bsnm{Traversa}, \binits{F.L.}},
\bauthor{\bsnm{Di~Ventra}, \binits{M.}},
\bauthor{\bsnm{Bonani}, \binits{F.}}:
\batitle{Generalized floquet theory: Application to dynamical systems with
  memory and bloch’s theorem for nonlocal potentials}.
\bjtitle{Phys.~Rev.~Lett.}
\bvolume{110}(\bissue{17}),
\bfpage{170602}
(\byear{2013})
\end{barticle}
\endbibitem

\bibitem[\protect\citeauthoryear{Magazz{\`u}
  et~al.}{2017}]{magazzu2017asymptotic}
\begin{barticle}
\bauthor{\bsnm{Magazz{\`u}}, \binits{L.}},
\bauthor{\bsnm{Denisov}, \binits{S.}},
\bauthor{\bsnm{H{\"a}nggi}, \binits{P.}}:
\batitle{Asymptotic floquet states of non-markovian systems}.
\bjtitle{Phys.~Rev.~A}
\bvolume{96}(\bissue{4}),
\bfpage{042103}
(\byear{2017})
\end{barticle}
\endbibitem

\bibitem[\protect\citeauthoryear{Magazz{\`u}
  et~al.}{2018}]{magazzu2018asymptotic}
\begin{barticle}
\bauthor{\bsnm{Magazz{\`u}}, \binits{L.}},
\bauthor{\bsnm{Denisov}, \binits{S.}},
\bauthor{\bsnm{H{\"a}nggi}, \binits{P.}}:
\batitle{Asymptotic floquet states of a periodically driven spin-boson system
  in the nonperturbative coupling regime}.
\bjtitle{Phys.~Rev.~E}
\bvolume{98}(\bissue{2}),
\bfpage{022111}
(\byear{2018})
\end{barticle}
\endbibitem

\bibitem[\protect\citeauthoryear{Trushechkin}{2019}]{trushechkin2019higher}
\begin{barticle}
\bauthor{\bsnm{Trushechkin}, \binits{A.}}:
\batitle{Higher-order corrections to the redfield equation with respect to the
  system-bath coupling based on the hierarchical equations of motion}.
\bjtitle{Lobachevskii J. Math.}
\bvolume{40}(\bissue{10}),
\bfpage{1606}--\blpage{1618}
(\byear{2019})
\end{barticle}
\endbibitem

\bibitem[\protect\citeauthoryear{Dekekr}{1987}]{dekker86}
\begin{barticle}
\bauthor{\bsnm{Dekekr}, \binits{H.}}:
\batitle{Noninteracting-blip approximation for a two-level system coupled to a
  heat bath}.
\bjtitle{Phys.~Rev.~A}
\bvolume{35},
\bfpage{1436}
(\byear{1987})
\end{barticle}
\endbibitem

\end{thebibliography}

\end{document}